\documentclass[showkeys]{revtex4}
\usepackage{psfig}

\newcounter{qcounter}

\begin{document}

\title{COHERENT PIONS FROM NEUTRINO SCATTERING OFF NUCLEI}

\author{M. VALVERDE}

\affiliation{Research Center for Nuclear Physics RCNP, Osaka University,\\
Ibaraki, 567-0047, Japan\\
E-mail: valverde@rcnp.osaka-u.ac.jp}

\author{J. E. AMARO}

\affiliation{Dept. de F{\'\i}sica At\'omica, Molecular y
Nuclear, Universidad de Granada\\ Granada, E-18071, Spain} 

\author{E. HERN\'ANDEZ} 

\affiliation{Grupo de F\'\i sica Nuclear, Dept.  de F\'\i sica
  Fundamental e IUFFyM \\ Facultad de Ciencias, Univ. de
  Salamanca\\ Salamanca, E-37008, Spain.}

\author{J. NIEVES}

\affiliation{Instituto de F\'\i sica Corpuscular (IFIC), Centro Mixto
  CSIC-Universidad de Valencia, Institutos de Investigaci\'on de
  Paterna\\ Valencia, Aptd. 22085, E-46071, Spain}

\author{M.~J. VICENTE VACAS}

\affiliation{Dpto. de F\'\i sica Te\'orica and IFIC, Centro Mixto
  CSIC-Universidad de Valencia, Institutos de Investigaci\'on de
  Paterna\\ Valencia, Aptd. 22085, E-46071, Spain}

\begin{abstract}
We describe a model for pion production off nucleons and coherent
pions from nuclei induced by neutrinos in the $1\,\text{GeV}$ energy
regime. Besides the dominant $\Delta$ pole contribution, it takes
into account the effect of background terms required by chiral
symmetry. Moreover, the model uses a reduced nucleon-to-$\Delta$
resonance axial coupling, which leads to coherent pion production
cross sections around a factor two smaller than most of the previous
theoretical estimates. Nuclear effects like medium corrections on
the $\Delta$ propagator and final pion distortion are included. 
\end{abstract}


\maketitle

\section{Introduction}\label{sec1:intro}

Neutrinos have been in the forefront of research in particle and
nuclear physics for a long time. One of these fields is the study of
pion production off nuclei induced by neutrinos. A proper
understanding of this process is necessary in the analysis of the
present generation of precision neutrino oscillation experiments.  For
instance, the $\pi^0$ produced in neutral currents (NC) is the most
important $\nu_\mu$-induced background to experiments like
MiniBoone\cite{AguilarArevalo:2007it} that are trying to measure
$\nu_\mu\to\nu_e$ oscillations in the neutrino energy range around
$1\,\text{GeV}$. Also of importance is the background that appears
from $\pi^+$ charged current (CC) production in $\nu_\mu\to \nu_x$
disappearance searches like T2K\cite{Hiraide:2006zq}.  Moreover the
pion is strongly coupled to the $\Delta(1232)$ resonance, and neutrino
scattering is presently the best way to access to the axial
nucleon-$\Delta$ transition couplings. The most complete information
in this regard comes from the bubble chamber data of
ANL\cite{Barish:1978pj,Radecky:1981fn} and
BNL\cite{Kitagaki:1986ct,Kitagaki:1990vs} where the target was cooled
deuterium.  However, in present oscillation experiments the target for
neutrino interaction is finite nuclei, for instance $^{12}$C (mineral
oil in MiniBoone) or $^{16}$O (water target in T2K). This introduces
sizable many-body effects that are difficult to disentangle from the
genuine single nucleon response to the neutrino probe.

In Sec.~\ref{sec:singlepion} of these paper we describe a
phenomenological model\cite{Hernandez:2007qq} for pion production
induced by neutrino scattering off free nucleons. This model takes
into account non-resonant background processes that are usually
neglected. These non-resonant processes are determined by chiral
symmetry and thus do not introduce free parameters. We then perform a
fit of the axial $N\text{-}\Delta$ parameters and discuss a possible
violation of the non-diagonal Goldberger-Treiman relation.  In
Sec.~\ref{sec:coherent} we extend the model to describe the CC
coherent process in which the final nucleus is left in its ground
state.  We will further try to discuss how the coherent reaction can
show some light on the values of the axial $N\text{-}\Delta$
couplings.  We shall focus on the CC process, but the model can be
easily extended to antineutrino reactions and NC processes.  The
interested reader could find further details in
Refs.~\cite{Hernandez:2007qq,Hernandez:2007ej,Amaro:2008hd,Hernandez:2009vm,Hernandez:2010bx}

\section{Single Nucleon Pion Production}\label{sec:singlepion}

\begin{figure}
  \begin{center}
    \psfig{file=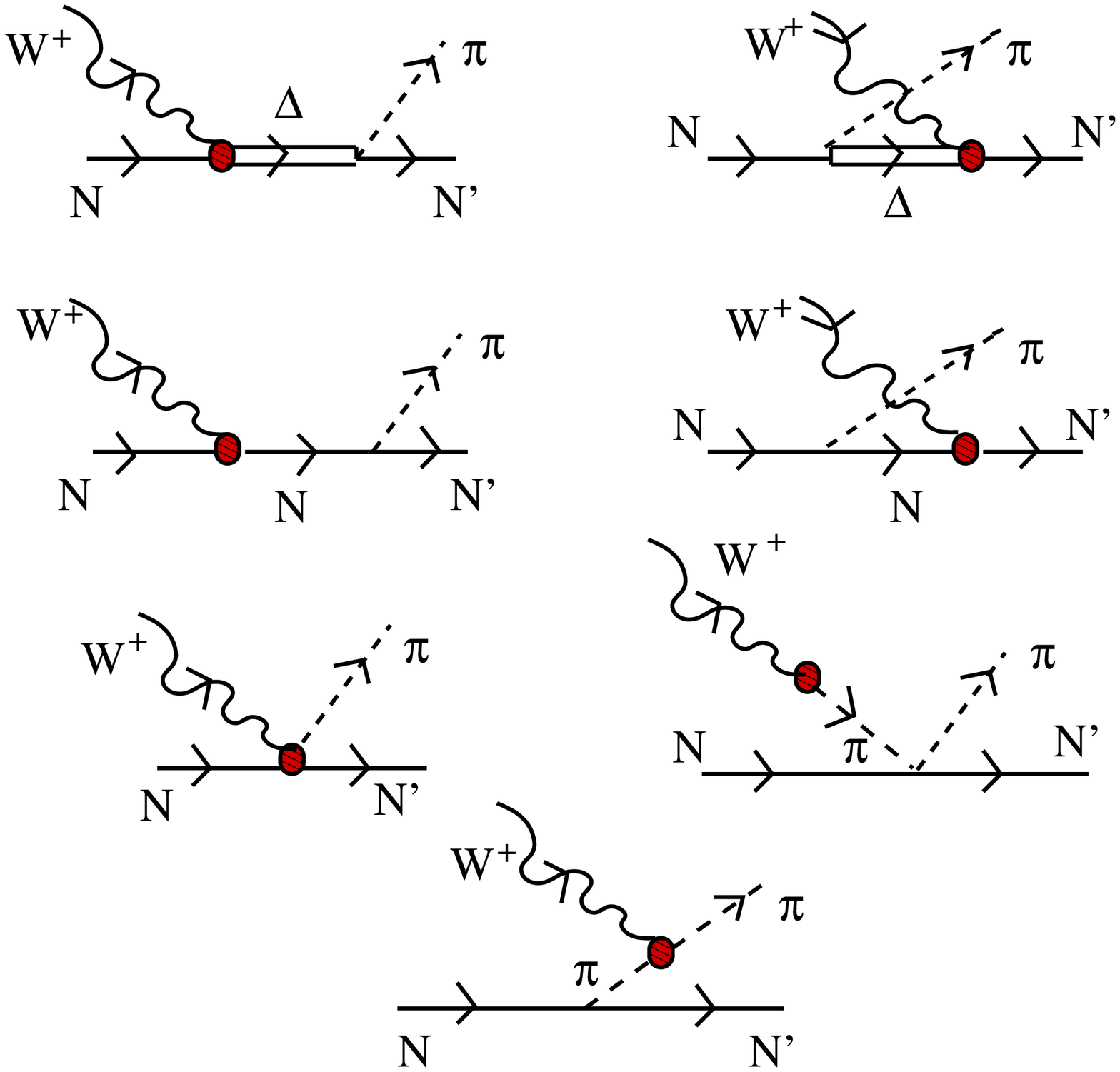,width=3in}
  \end{center}
  \caption{Set of diagrams for the model for the $W^+N\to N^\prime\pi$
    reaction.}
  \label{fig:diagrams}
\end{figure}

Here we review the model for the free nucleon reaction
\begin{equation}
  \nu_l (k) +\, N \to l^- (k^\prime) + N^\prime +\, \pi^+(k_\pi)
\label{eq:freereac}
\end{equation}
as introduced in Ref.~\cite{Hernandez:2007qq}. This model
considers the dominant $\Delta$ pole mechanism in which the neutrino
excites a $\Delta(1232)$ resonance that subsequently decays into
$N\pi$. In our model we have also included non-resonant background
terms as required by chiral symmetry, see
Fig.~\ref{fig:diagrams}. Some previous
works\cite{Fogli:1979cz,Fogli:1979qj,Sato:2003rq} also considered
background terms, though they were not consistent with the chiral
counting.

The vector part of the interaction $N\text{-}\Delta$ can be related to
the electromagnetic current by imposing conservation of the vector
current. For photon induced reactions extensive experimental data
exist and in Ref.~\cite{Lalakulich:2006sw} they were employed to
fit the vector current couplings. We shall use this fit in our work.
Unfortunately the axial $N\text{-}\Delta$ current is not so well
studied. The usual approach is to parameterize the interaction in terms
of four form factors $C^A_{3,4,5,6}(q^2)$.  One can assume partial
conservation of the axial current (PCAC) and obtain the relation
$C^A_6 = C^A_5 M^2/(m_\pi^2-q^2)$, with $M$ the nucleon
mass. Furthermore, one can deduce\cite{Adler:1964yx} from dispersion
relations the following conditions: $C_3^A(q^2) = 0$ and $C^A_4 =
-C^A_5/4$. Thus we are left with only a free form factor, the dominant
one $C_5^A(q^2)$. A different number of parameterizations have been
proposed for this form factor, nevertheless the experimental data are
quite limited and thus a simple dipole form
\begin{equation}
C^A_5(q^2) = \frac{C^A_5(0)}{\left(1-q^2/M_A^2\right)^2}
\end{equation}
should be enough. In order to keep the axial transition radius in the
range $0.7\text{--}0.8\,\text{fm}$ one expect the axial mass to have a
value of around $M_A \sim 0.85\text{--}1.0\,\text{GeV}$.  Furthermore
one can assume the well known Goldberger-Treiman relation (GTR) for
$\pi NN$ coupling to be also valid for the $\pi N\Delta$ coupling and
thus obtain
\begin{equation}
C^A_5(0) = \sqrt{\frac{2}{3}}f_\pi\frac{f^*}{m_\pi} = 1.2 \, ,
\label{eq:gtr}
\end{equation}
where $f_\pi=93\,\text{MeV}$ is the pion decay constant and $f^*=2.2$,
the $\pi N\Delta$ coupling. Unfortunately, there are no constraints
from Chiral Perturbation Theory\cite{Geng:2008bm} and the lattice QCD
calculations\cite{Alexandrou:2006mc} are still inconclusive.

Most of the approaches in the literature assume $\Delta$ dominance,
that is, only include the first two diagrams in
Fig.~\ref{fig:diagrams}. We improve this situation by including
non-resonant contributions\cite{Hernandez:2007qq} required by chiral
symmetry. In addition to the $\Delta(1232)$ pole ($\Delta P$) (first
row) mechanism the model includes background terms required by chiral
symmetry: nucleon (second row) pole terms contact and pion pole
contribution (third row) and pion-in-flight term.  We calculate them
by using the SU(2) non-linear $\sigma$ model Lagrangian.
The only parameters in the theory are the pion and nucleon masses and
the pion decay constant. All other couplings are completely fixed by
the theory, so no new parameter is introduced.

\begin{figure}[tbh]
  \begin{center}
    \makebox{    \psfig{file=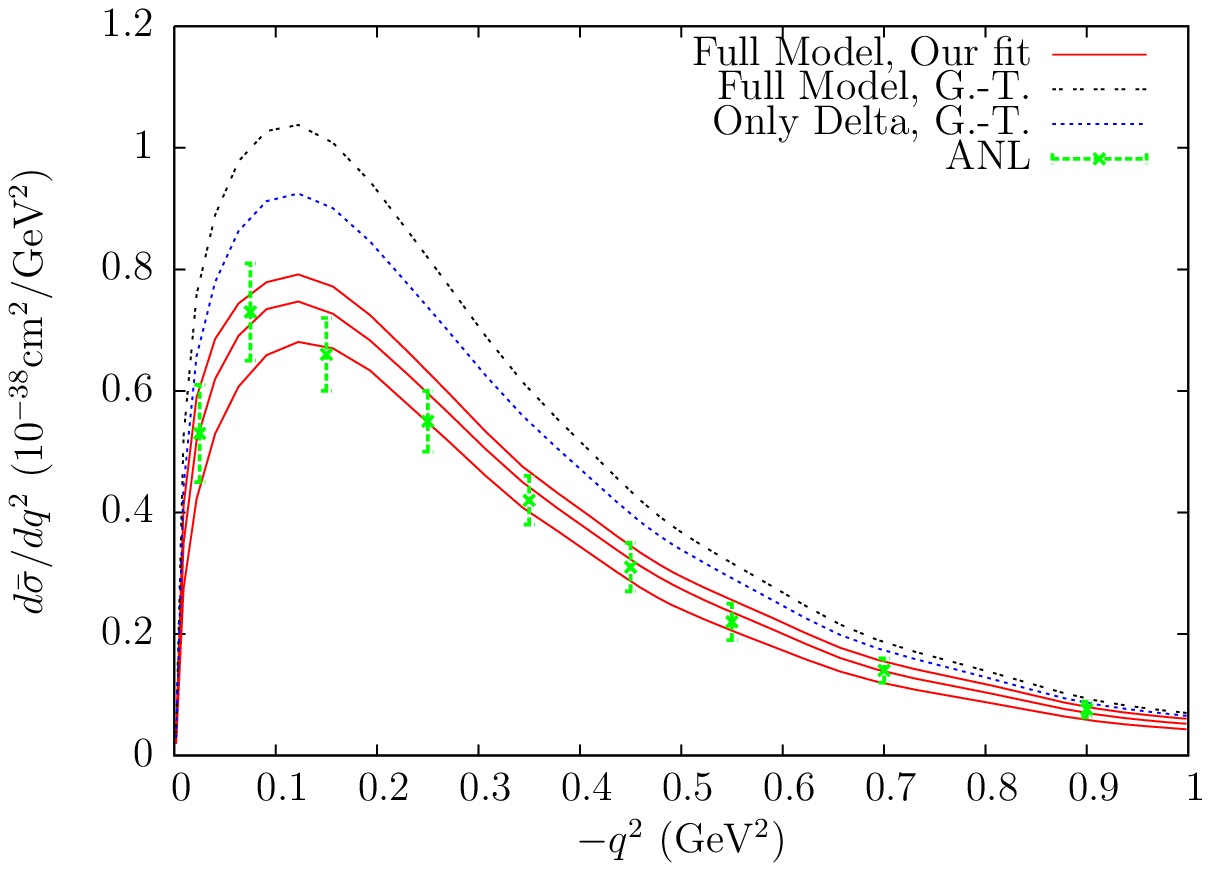,width=2.15in} \psfig{file=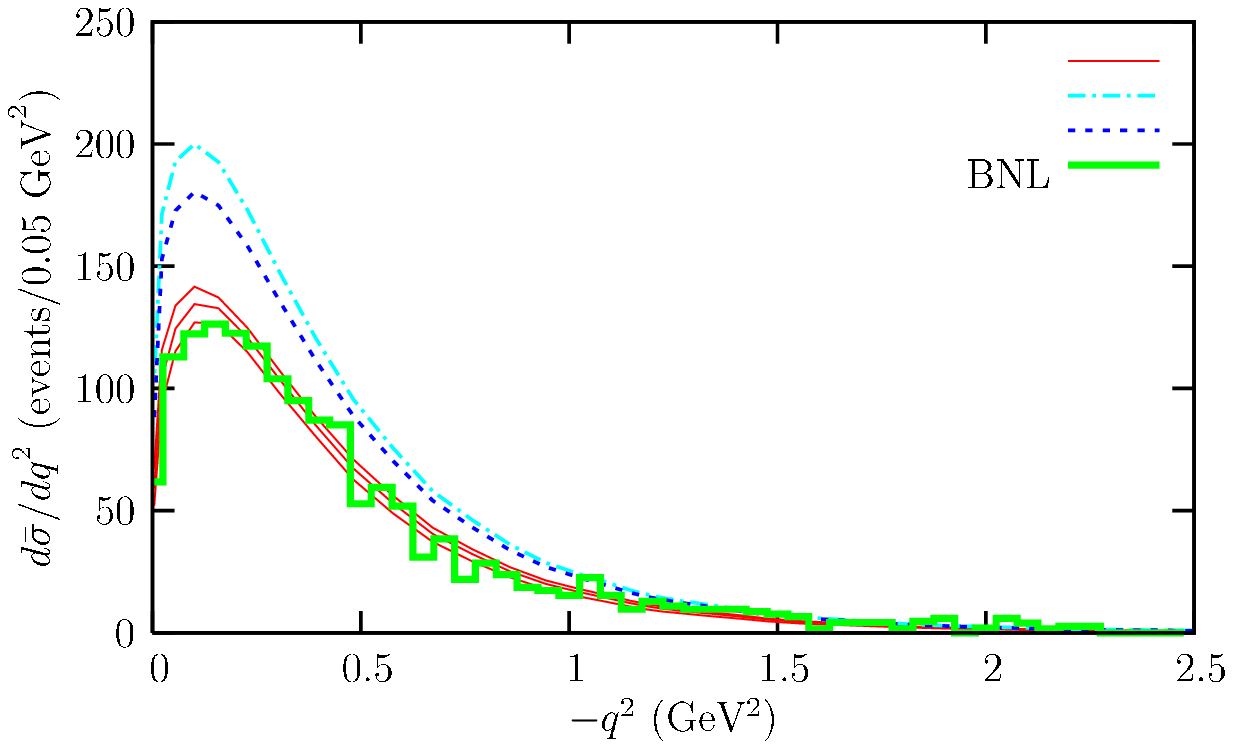,width=2.25in}}
  \end{center}
  \caption{\footnotesize Flux averaged $q^2-$differential $\nu_\mu p
    \to \mu^- p \pi^+$ cross section for the ANL (left) and
    BNL (right). Dashed lines stand for the contribution
    of the $\Delta P$ mechanism with the GTR assumption for
    $C_5^A$. We also plot results with the full model of
    Fig.~\ref{fig:diagrams}, assuming GTR (dashed-dotted) and with our
    best fit parameters, Eq.~(\protect\ref{eq:besfit}).}
  \label{fig:anl-bnlq2}
\end{figure}

We found (see Fig.~\ref{fig:anl-bnlq2}) that these background terms produced significant effects in
all channels, namely an enhancement of about $10$\% in the cross section
that resulted in a disagreement with the ANL data.
As a result we had to readjust the strength of the dominant $\Delta$
pole contribution.  The least known ingredients of the model are the
axial nucleon-to-$\Delta$ transition form factors, of which $C_5^A$
gives the largest contribution. This strongly suggested a refit of
that form factor to the experimental data, which we did by fitting the
flux-averaged $\nu_\mu p\to \mu^- p \pi^+$ ANL $q^2$-differential
cross section for pion-nucleon invariant masses\footnote{This cut was
  introduced in order to avoid the effects of resonances higher than
  the $\Delta(1232)$.} $W < 1.4\,\text{GeV}$.  The obtained parameters
were
\begin{equation}
  C^A_5(0) = 0.87 \pm 0.08 \, , \quad M_A = 0.985 \pm 0.082 \,\text{GeV}
  \label{eq:besfit}
\end{equation}
with a $\chi^2/\text{dof}=0.4$ and a correlation coefficient
$r=-0.85$, that amounts to a $30$\% reduction of the GTR prediction.
Thus, our full model leads to an overall better description of the
data for one-pion production reactions off the nucleon. This reduction
of the $C_5^A(0)$ value is consistent with recent results in lattice
QCD\cite{Alexandrou:2006mc} and quark
models\cite{BarquillaCano:2007yk}.

Recently other fits have been proposed. For instance, in
Ref.~\cite{Leitner:2008ue} they keep the GTR but introduce a
non-dipole form factor with additional parameters. As in the ANL data
the relevant phase space is around $q^2=0.1\,\text{GeV}^2$, they could
keep the GTR at the cost of having a large dependence on $q^2$ for the
form factor, that yields a large, somehow unphysical, axial transition
radius of around $1.4\,\text{fm}$. Furthermore neither statistical
errors nor correlation factors are given in that reference. Another
analysis\cite{Graczyk:2009qm} raised new questions, namely the effect
of deuterium wave function on the cross section and the flux
uncertainties in the ANL and BNL data. The authors of this latter work
took into account both effects, though we believe that their
statistical analysis is not quite robust (see discussion in
Ref.~\cite{Hernandez:2010bx}). Furthermore they only took into
account the dominant $\Delta$ contribution. The inclusion of deuterium
wave function reduces the cross section about an $8$\%, so this
somehow compensates the neglect of the non-resonant background, and
they obtained a best fit of $C_5^A(0) = 1.19\pm 0.08$ in agreement
with the GTR assumption.

Recently\cite{Hernandez:2010bx} we have improved our fit of
Ref.~\cite{Hernandez:2007qq}, improving the lines suggested in
Graczyk et al. In summary in this new fit
\begin{list}{\arabic{qcounter}:~}{\usecounter{qcounter}}
\item all diagrams in Fig.~\ref{fig:diagrams} are included;
\item the fitted data are the full ANL data set and the BNL total
  cross sections at the three lowest neutrino energies (we neglect
  higher energies where the effect of higher resonances beyond the
  $\Delta(1232)$ must be addressed; the BNL $q^2$-differential cross sections were not taken into account as they are not normalized);
\item deuterium wave function effects were introduced following the prescription of Ref.~\cite{AlvarezRuso:1998hi};
\item the form factors $C_3^A$ and $C_4^4$ were tentatively included
  in one fit, though the data were found to be quite insensitive to
  their values so we decided to stick to the Adler's assumption; and
\item the uncertainties in ANL and BNL flux normalization are introduced as fully correlated systematic errors.
\end{list}
In this way we obtain a best fit of $C_5^A(0) = 1.00\pm 0.11$ and $M_A
= 0.93\pm 0.07\,\text{GeV}$ with a goodness of fit value of
$\chi^2\text{d.o.f.} = 0.42$. Thus we observe a violations of the off
diagonal Goldberger-Treiman relation at the level of $2\sigma$.

\section{The Coherent Reaction}\label{sec:coherent}

Here we describe our model\cite{Amaro:2008hd} for the coherent reaction
\begin{equation}
  \nu_l (k) +\, A_Z|_{gs}(p_A) \to l^- (k^\prime) + A_Z|_{gs}(p^\prime_A) +\, \pi^+(k_\pi)
\label{eq:reac}
\end{equation}
where the target nucleus $A_Z$ is left in the ground state ($gs$). To
calculate the amplitude of this process we sum over all individual
nucleon wave functions, which are modelled by a Fermi gas in local
density approximation. The individual nucleon amplitudes are modelled
following the model of the previous section, using the fit of
Eq.~\ref{eq:besfit} for the $C_5^A$ form factor. On top of that a
number of many-body effects are introduced. In first place we take
into account the in-medium modifications\cite{Oset:1987re} of the
$\Delta(1232)$ properties. This implies a shift in the pole mass
towards lower energies and most importantly a net broadening of the
width (the opening of new decay channels in the nuclear medium
compensates the Pauli blocking of the $\pi N$ decay channel). Also
important is the distortion of the outgoing pion by strong interaction
with the nucleus. Thus we consider the wave function of the pion to be
the outgoing solution to the Klein-Gordon equation with a microscopic
optical potential\cite{Nieves:1991ye} whose imaginary part takes into
account the inelastic interactions of the pion with the nucleus, that
thus disappear from the coherent channel.  We must emphasize here that
solving the Klein-Gordon equation is the correct way of describing the
distortion of the outgoing pion.  Other approaches use either a Monte
Carlo simulation\cite{AguilarArevalo:2008xs} or include an attenuation
factor fitted to the pion nuclei scattering cross
section\cite{Nakamura:2009iq}.  The first procedure, though physically
sound, can be a bit misleading as it includes in its cross section
processes (like quasi-elastic scattered pions) that do not leave the
nucleus in it ground state, thus are not coherent. This kind of models
are used in the analysis of MiniBoone experiment, thus making a bit
messy the direct comparison between theoretical models and
experimental results (see discussion in Ref.~\cite{Amaro:2008hd}).
The second approach is an oversimplification, as the pion-nucleus
scattering is quite a different process from the neutrino induced pion
production. As pion nucleus interaction is governed by the strong
interaction, the incoming pion interacts strongly with the nuclear
surface, thus the pion nucleus cross section is quite insensitive to
the details of the nuclear core. On the other hand, neutrino
scattering is a weak process, dominated by nuclear density, so pions
are mostly produced in the deep, high density regions of the
nucleus. The physics of pion interaction is thus quite different in
pion scattering off nuclei and pion production by electroweak probes.

\begin{figure}
\begin{center}
  \makebox{\psfig{file=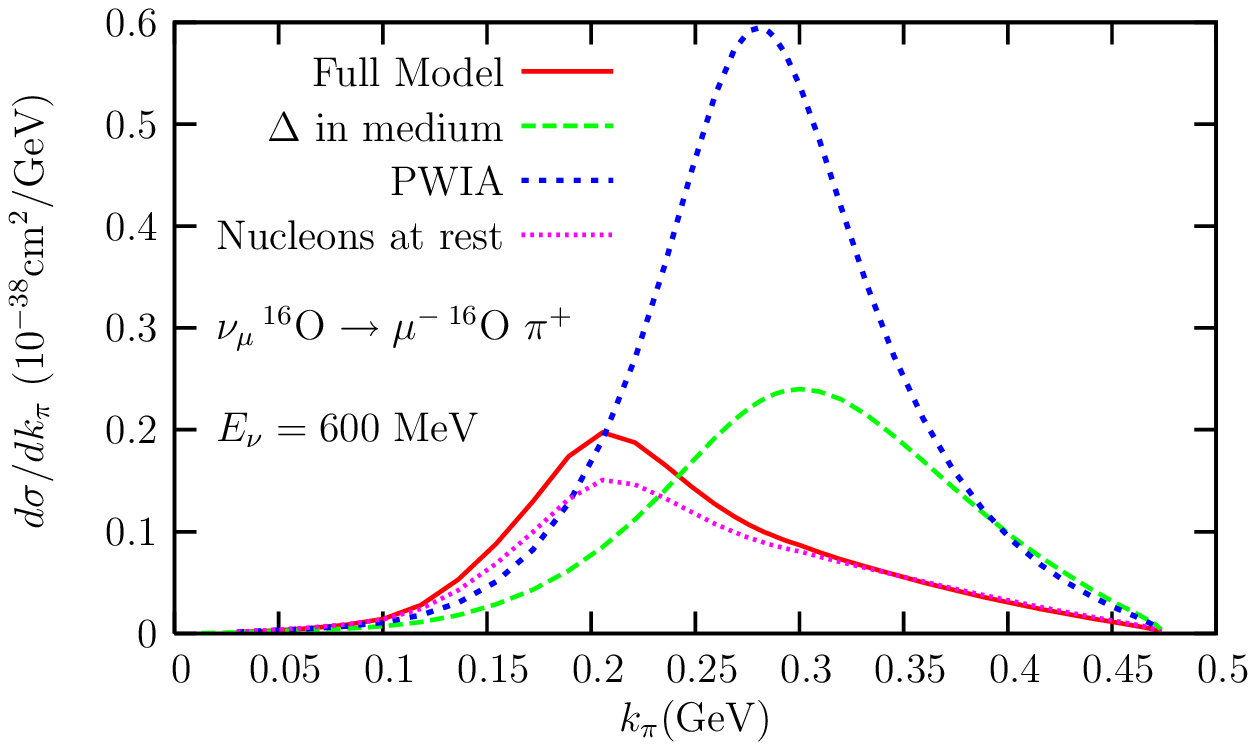,width=2.25in}
    \psfig{file=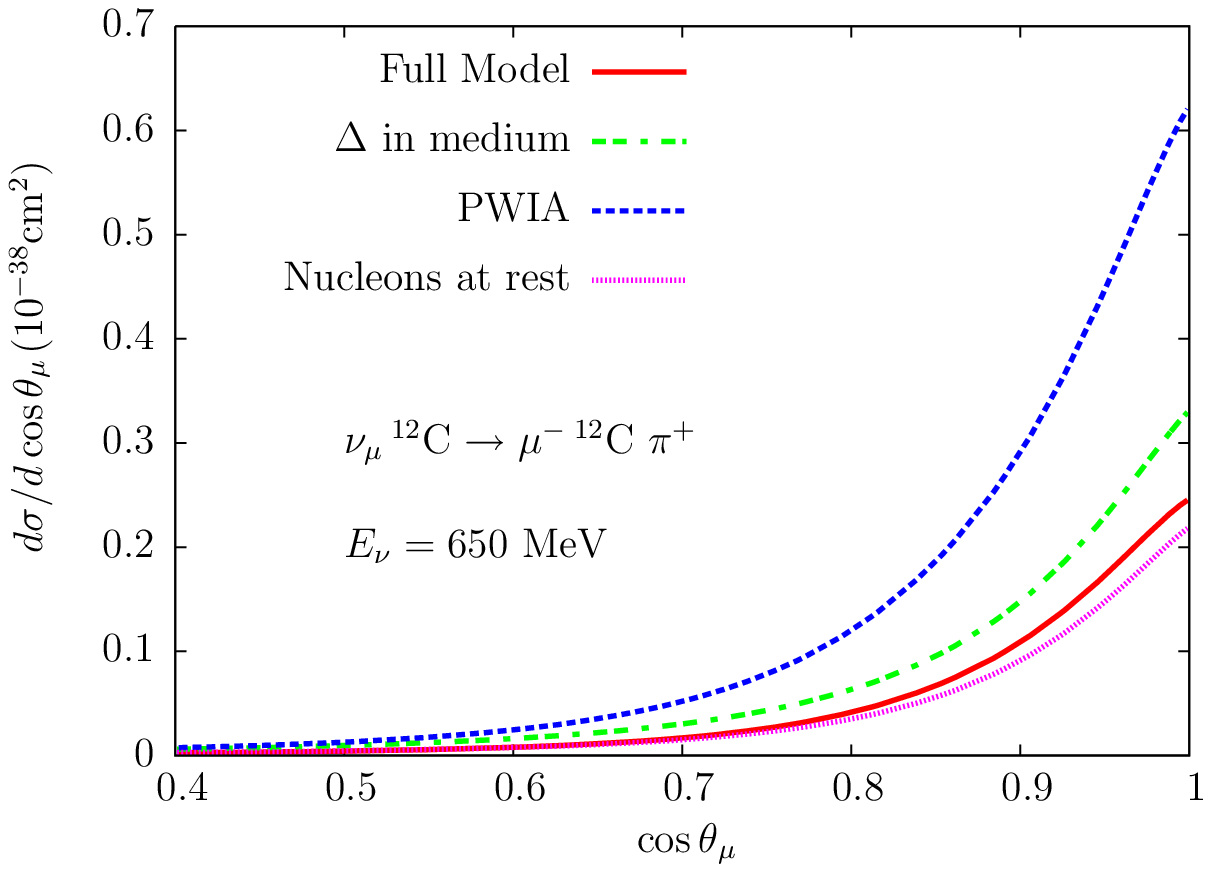,width=2.25in}}
\end{center}
\caption{Pion momentum (right) and angular (left) differential cross section.}
\label{fig:dspion}
\end{figure}

In left panel of Fig.~\ref{fig:dspion} we show the pion momentum
distribution for CC coherent pion production, in the peak energy
region of the T2K experiment. 
Including $\Delta$ in-medium self-energy (long-dashed line) reduces
the PWIA results (short-dashed line).  Further inclusion of pion
distortion (full model, solid line) reduces the cross section, and the
peak is shifted towards lower energies, reflecting the strong
absorption and the higher probability of a quasi-elastic collisions of
the pion in the $\Delta$ kinematical region.  The total cross section
reduction is around $60\%$.  Similar nuclear effects were already
studied in
Refs.~\cite{AlvarezRuso:2007tt,AlvarezRuso:2007it}. However, the
authors of these references neglected the nucleon momenta in the Dirac
spinors. The effect of this approximation (nucleons at rest, dotted
line) results in a $\sim 15$\% decrease of the total cross section.
In the right panel of Fig.~\ref{fig:dspion} we show the pion angular
distribution with respect to the incoming neutrino direction. The
reaction is very forward peaked, as expected due to the nucleus form
factor.  The angular distribution profile keeps its forward peaked
behavior after introduction of nuclear medium effects. Furthermore we
corrected some numerical errors in the mentioned papers. However one
must be aware that this model does not take into account the
non-localities\cite{Leitner:2009ph} in the $\Delta$ propagation.  We
believe this effect is partially taken into account in an effective
fashion by our treatment of the $\Delta$ in nuclear medium;
nevertheless further studies would be interesting.

\begin{table}

{\begin{tabular}{@{}cccc@{}}\toprule
Reaction                 & Experiment &$\sigma (10^{-40}\,\text{cm}^2)$& $\sigma (10^{-40}\,\text{cm}^2)$ Exp\\ 
\colrule
CC $\nu_\mu + ^{12}$C    & K2K        &        & $<7.7$            \\

NC $\nu_\mu + ^{12}$C    & MiniBoone  & $3.33$ & $7.7\pm1.6\pm3.6$ \\
CC $\nu_\mu + ^{12}$C    & MiniBoone  & $4.46$ &               \\

CC $\nu_\mu + ^{16}$O    & T2K        & $4.19$ &               \\
CC $\nu_\mu + ^{12}$C    & T2K        & $3.54$ &               \\
\botrule
\end{tabular}
\caption{\footnotesize Total cross sections for the coherent process.
We neglect the highest $10\%$ of the energy spectrum.}
}
\label{tab:res}
\end{table}
In Table~\ref{tab:res} we compare our model with present results of
K2K experiment\cite{Hasegawa:2005td} and show our predicitions for
MiniBoone and T2K.  Our prediction, subject to sizable
uncertainties, lies well below the K2K upper bound, due to the use of
a low value for $C_5^A(0)$, while our prediction for the $\nu_\mu$ NC
MiniBoone cross section is notably smaller than that given in the PhD
thesis of J.~L.~Raaf\cite{Raaf:2005up}. However, we believe (see discussion in
Refs.~\cite{Amaro:2008hd,Hernandez:2009vm} that the MiniBoone
analysis might importantly overestimate this cross section, not only
because some of the $\pi^0$s which undergo FSI collisions are
accounted for instead of being removed, but also because a possible
mis-match between the absolute normalisation of the background and
coherent yields. Note that the K2K and MiniBoone results seems somehow
incompatible with the approximate relation $\sigma_{\text{CC}} \approx
2\sigma_{\text{NC}}$, which would be expected from $\Delta$ dominance
and neglecting finite muon mass effects.

\section*{Acknowledgments}
M. V. acknowledges a Postdoctoral Fellowship from the Japanese Society
for the Promotion of Science (JSPS). Research supported by DGI
contracts FIS2008-01143, FIS2006-03438, FPA2007-65748 and
CSD2007-00042, JCyL contracts SA016A07 and GR12, Generalitat
Valenciana contract PROMETEO/2009/0090 and by EU HadronPhysics2
contract 227431.

\end{document}